\documentclass{article}
\usepackage{arxiv}
\usepackage[utf8]{inputenc} 
\usepackage[T1]{fontenc}    
\usepackage{hyperref}       
\usepackage{url}            
\usepackage{booktabs}       
\usepackage{amsfonts}       
\usepackage{nicefrac}       
\usepackage{microtype}      
\usepackage{lipsum}		
\usepackage{graphicx}
\usepackage{doi}
\usepackage{color}
\def\etal {{\em et al.}}

\title{Overcoming 1 part in $10^9$  of Earth angular rotation rate measurement with the G Wettzell data}

\author{ A. D. V. Di Virgilio and G. Terreni\\
	INFN-Pisa largo B. Pontecorvo 3, 56127 Pisa, Italy\\
	\texttt{angela.divirgilio@pi.infn.it} \\
	\And
	A. Basti, N. Beverini, G. Carelli, D. Ciampini, F. Fuso, E. Maccioni and P. Marsili\\
	Dep.of Physics Univ. of Pisa, largo B. Pontecorvo 3, 56127 Pisa, Italy\\
	\And 
	J. Kodet and K. U. Schreiber \\
	Technical University of Munich, \\ FESG, Geodetic Observatory Wettzell, \\93444 Bad Kotzting, Germany\\
}

\renewcommand{\shorttitle}{\textit{arXiv} Template}

\hypersetup{
pdftitle={A template for the arxiv style},
pdfsubject={q-bio.NC, q-bio.QM},
pdfauthor={David S.~Hippocampus, Elias D.~Striatum},
pdfkeywords={First keyword, Second keyword, More},
}

\begin{document}
\maketitle
\begin{abstract}
 The absolute measurement of the Earth angular rotation rate  with ground-based instruments becomes challenging if the 1 part in $10^9$ of precision has to be obtained. This threshold is important for fundamental physics and for geodesy, to investigate effects of General Relativity and Lorentz violation in the gravity sector and to provide the fast variation of the Earth rotation rate. 
 High sensitivity Ring Laser Gyroscopes (RLG) are currently the only promising technique  to achieve this task in the near future, but their precision has been so far limited by systematics related to the laser operation. 
 In this paper we analyze two different sets of observations, each of them three days long. They were obtained from the G ring laser at the Geodetic Observatory Wettzell. The applied method has been developed for the GINGERINO ring laser in order to identify and extract the laser systematics.  For the available data sets the residuals show mostly white noise behavior and the Allan deviation drops below 1 part in $10^9$ after an integration time of about $10^4$~s. 

\end{abstract}

\section{Introduction}
At present large scale ring laser gyroscopes (RLGs) are the most sensitive instruments to measure absolute angular rotation rates\cite{uno, due, tre, EPJC21, ER1, HUST}. They are based on high finesse square optical cavities where an active medium is present, and two counterpropagating laser beams are generated. The frequencies of the two beams depend on  the effective time the photon takes to travel along the perimeter, which  is different when the gyro is rotated. However, non-reciprocal effects in the cavity from the laser excitation process will cause a systematic bias and this has to be avoided.  
 
 Furthermore, there can be small deviations from the expected rate of rotation, due to fundamental laws of physics. One example is the precession of the frame of reference, as predicted by General Relativity (GR), related to gravitoelectric and gravitomagnetic effects. 

For a RLG rigidly connected with the Earth surface, the observed Sagnac frequency is the difference in angular frequency of the two laser beams $\omega_s$:
\begin{equation}
  \omega_s =8 \pi \frac{A}{P\lambda}\Omega_\oplus\cos\theta\;, 
\end{equation}
where $A$ is the area and $P$ the perimeter of the ring cavity, $\lambda$ the optical wavelength, $\Omega_\oplus$ the angular rotation rate, equivalent to the Earth rotation rate. $\theta$ is the angle between the optical cavity area vector and the $\Omega_\oplus$ rotation axis.

 G has demonstrated a sensitivity of 12 prad/s in 1 s integration time and is able to operate continuously and unattended for months.\cite{please_Ulli} The sensitivity is a function of the size of the ring cavity and cavities of 77 and 121 m perimeter,\cite{NZ1} have been explored, several years ago. More recently the four component ring laser array ROMY has been installed in the geophysical observatory of Bavaria, Germany, which comprises 4 RLGs with triangular cavities, each with a perimeter of 36 m \cite{due, romy2}. Typically, the most sensitive existing devices capable of long term continuous operation employ square optical cavities with several meters on a side, rigidly attached to the ground. The mechanical structure of the optical cavity plays a big role in the performance of RLG and monolithic optical cavity structures have been the first RLGs to obtain relevant performance, getting close to  tens prad/s level of sensitivity \cite{uno}. The G RLG, in operation at the Geodetic Observatory Wettzell, is the best known example and the optimal choice to have a very stable laser cavity. However, it is very difficult to build and cannot be reorientated easily. Over more than 20 years, heterolythic (HL) cavities have been developed, basically consisting of a rigid mounting frame giving hold to different mechanical elements, which support the mirrors contained inside metallic vacuum chambers. 

GINGERINO has been the first HL RLG operating continuously with high sensitivity \cite{G90}. GINGERINO takes advantage of the quiet location inside the Gran Sasso, Italy, underground laboratory. Despite that, the standard deviation of the reconstructed Sagnac frequency of GINGERINO is typically more than 50 times larger than that of G \cite{tre,EPJC21}, mostly due to small mechanical issues, since the composed structure is not yet rigid enough\cite{EPJP2020}. Work is currently in progress to improve the mechanical HL design in order to increase the sensor stability. Despite these mechanical shortcomings in GINGERINO, the previous investigations \cite{tre,EPJC21} indicate that the intrinic noise level is lower than the expected shot noise limit for a RLG\cite{ chow}. This model does not take the presence of the active medium inside the cavity into account and there is work in progress to improve the theoretical model by the Italian group of GINGER.

The main signal obtained from a RLG is the interferogram of the two counterpropagating beams transmitted at one corner of the square cavity. The Sagnac frequency $\omega_s$, which is caused by the rate of rotation, must be not confused with the actually observed beat note $\omega_m$ of the interferogram \cite{Lamb, Aronowitz,Beghi,Cuccato}, since the latter is biased by laser systematics, usually caused by backscatter coupling and a null shift contribution. An original analysis scheme to remove laser systematics has been developed and successfully applied to analyse the data of GINGERINO  \cite{DiVirgilio2019,DiVirgilio2020,tre,EPJC21}. This effort shows how any change due to the laser dynamics can be approximated with the sum of several terms, analytically evaluated using the laser parameters of the Lamb theory. These in turn can be determined from the signals provided by the RLG. It is important to remark that these laser systematics are not a stochastic effect, since they are caused by all sorts of small changes affecting the active optical cavity. These effects are much smaller in monolithic RLGs.

\section{Analysis procedure}
The laser dynamics can be described by a set of differential equations containing several parameters, which are known as Lamb parameters \cite{Lamb,Aronowitz,DiVirgilio2019, DiVirgilio2020}. They can be calculated from the diagnostic signals, available from the ring laser intensities, in particular those providing the AC ($IS_{1,2}$) and the DC ($PH_{1,2}$) levels of the detectors, each looking at a different laser beam (hereafter denoted as monobeam signals) and their relative  phase offset $\epsilon$.  
The model refers to the intracavity power, while the signals are taken outside the cavity; it is therefore necessary to know the values of mirror transmission and detector gain.

The analysis proceeds in steps. The first approximation of the Sagnac frequency, denoted as $\omega_{s0}$, is analytically evaluated as follows:
\begin{equation}
\omega_{s0} = \frac{1}{2} \sqrt{\frac{ 2  I_{S1}   I_{S2} \omega _m^2 \cos (2 \epsilon )}{ PH_{1}  PH_{2}}+\omega _m^2}+\frac{\omega _m}{2}\;,
\label{approx}
\end{equation}
where $\omega_m$ is the experimental interferometric angular frequency, that is the beat note of the interference of the two counterpropagating beams coming out from the cavity. It must be noted that the monobeam intensities enter Eq.~\ref{approx} as the ratio of their AC and DC components, leading to mirror transmission and the differences in electronic gains to play a smaller role. For many applications, $\omega_{s0}$ is indeed a good approximation of $\omega_s$.

Nevertheless, the different terms of the equation are certainly affected by errors, as for example the dark current of the applied photodiodes or any non-linearity in the electronic gain. In order to improve the $\omega_{s0}$ evaluation, six terms, namely $\omega_\xi$, have been elaborated, assuming small errors in the measurement data used in Eq.~\ref{approx} and expanding the equation to the first order. The first term was already  discussed in the first paper dedicated to the analysis method \cite{DiVirgilio2019}, the other ones have been added more recently. 
We remark that the six $\omega_\xi$ terms do not depend on the laser dynamics and are meant to improve $\omega_{s0}$ by correcting errors associated with the measurements themselves. 

The null shift depends on the Lamb parameters associated with the laser excitation and is obtained by the term $\omega_{ns1}$, the first order expansion of the theory. Higher order expansion terms can be calculated, but in the following analysis only the first one will be used. The null shift is strictly connected to the non-reciprocity of the optical path in the two directions, related in turn to dissipative processes. In terms of the Lamb parameters of the laser functions, $\omega_{ns1}$ is connected to the difference $\mu_c-\mu_{cc}$ between the cavity losses, where $\mu_{c,cc}$ represent losses in clockwise and counter-clockwise propagation directions. However, it has been demonstrated in  \cite{DiVirgilio2019} that, assuming a quasi-stationary laser and the same value for the parameter $\beta$ for the two beams\footnote{$\beta_{1,2}$  is the self-saturation parameter of  the laser transition for the two counter-propagating beams. In the model it is assumed $\beta_c = \beta_{cc}=\beta$.}, the loss difference is made explicit, so that  only one of the two $\mu_{c,cc}$ is a free parameter. In the following, we will consider only the clockwise cavity loss, which will be indicated simply as $\mu$;  $\omega_{ns1}$ and $\beta$ are proportional to $\mu$ and completely defined by the theory in terms of the available signals. The null shift correction has values in the region of mHz (several ppm), so the accuracy of $\mu$, usually measured by the ring-down time of the optical cavity with percent accuracy, can severely limit the $\omega_s$ reconstruction.
The analysis developed for GINGERINO assumes constant the plasma temperature and pressure, while $\mu$ changes with time,
accordingly any change is interpreted as change of $\mu$, but it could be due to the other parameters, or to the electronic circuit regulating the gain tube. In the future the analysis model will be further expanded in order to better identify the origin of the changes.
A suitable procedure is developed to take into account variations of $\mu$ with time and refine the identification of the null shift.   
 Changes of $\mu$ in time, however, are small and slow enough to not invalidate the assumption of a quasi-stationary regime. It is possible to consider $\mu(t) = \mu(t_0) + \delta\mu(t)$, $t_0$ being the origin of the expansion in the series. It is convenient to describe $\delta\mu(t)$ using the available signals. To this aim, the laser gain monitor signal can be used, but in more recent versions of the analysis the Lamb parameter $\beta$ has been considered, since it is rather constant in time
 and proportional to $\mu$.  By   using the relationship reported in the appendix of \cite{DiVirgilio2020} and the parameters of  the RLG G, we have
 \begin{equation}
 \beta =\frac{2.98772 \mu }{2.98772 - PH_1}\;,
 \label{eq:beta}
\end{equation}
where  $PH_1$ is the DC value of monobeam 1, in Volts.
Equation \ref{eq:beta} shows that $\mu$ is a proportionality constant, accordingly the quantity that can be evaluated does not contain $\mu$. Assuming $\mu(t) = \mu(t_0) + \delta\mu(t)$, and taking into account that $\beta$ is rather constant, since it is related to the laser transition,   $\delta\mu(t)$ can be evaluated  at the first approximation as follows:
\begin{eqnarray}
\bar{\beta}&=&\frac{\beta(t)}{\mu} \nonumber\\          
    \delta\mu(t) &=&\mu(t_0)\times\left(\frac{\bar{\beta}(t_0)}{\bar{\beta}(t)}-1\right)\;.
\end{eqnarray}
 In this way, $\mu(t_0)$, the loss at the time $t_0$,  has to be determined by statistical means, with $t_0$ arbitrarily chosen: in the present analysis $t_0$ is the central point of the data set.
The Lamb parameters can be considered constant over short time intervals; therefore, the stationary mathematical relationships can be considered valid for short time intervals. Accordingly the required analytical relationships are elaborated at high frequency rate, and decimated afterwards down to the desired low frequency rate, details can be found in the related literature\cite{DiVirgilio2019,DiVirgilio2020}.

The second step of the analysis determines $\omega_s$ with a linear regression to optimize the subtraction of the $\omega_\xi$ terms, $\omega_{ns1}$ and $\omega_{ns1}\times\delta\mu$. 
In this way, $\omega_s$ is recovered, but it is still necessary to further identify and subtract other known signals. The G RLG is well isolated, but in any case affected by global and local motion of the Earth crust, such as the diurnal polar motion and the solid Earth tides. In the present analysis the direct effect of the environmental disturbances on the mechanical apparatus is considered negligible, owing to the monolithic and stable ZERODUR\footnote{ZERODUR is a glass with very low thermal expansion coefficient, that can be considered practically negligible at room temperature.} structure. It is assumed that all external effects, hereafter denoted geodetic components, can be described by sufficiently accurate models. The geodetic components are well known, but they are added to the linear regression along with the other terms in order to avoid biases induced by previous analyses. The terms relevant to describe the main geodetic components are taken from model data provided by the G group. Furthermore, it is also possible to take the angular rotation around the vertical in the local reference frame at the G latitude and longitude into account. They are based on the publications of the International Earth Rotation and Reference Systems Service (IERS)\footnote{\texttt{https://hpiers.obspm.fr/eop-pc/index.php?index=C04\&lang=en}.} We denote relevant terms as $FGEO$. It is important to remark that $FGEO$ contains the Chandler and the Annual wobble, whose values are published with several days delay. For this reason these signals are not included in the near realtime data files of G. Finally, we have verified that the calibration and alignment of G is at the level of $0.3\%$.

\subsection{Details of data analysis}\label{details}

The whole analysis is based on linear regression (LR) \cite{Kay,Neter, Sen, direnzo}. Only terms with p-values below $0.4$ are kept in the linear regression, carried out by using the MATLAB routine $fitlm$, with the option $RobustOpt$ on. It has been checked that the final p-values are always below $0.2$.
The reconstruction of the AC and DC signals is based on Hilbert transform: data are band-pass filtered around the beat note (in a range $\pm12$ Hz around $\omega_m$) and Blackman windowing is implemented. This part of the analysis has been validated by processing a known ideal sinusoidal signal, without noise addition. It has been checked that  the Hilbert transform routine leads to recover the expected frequency within 7 nHz before decimation, and within 5 nHz after decimation down to the rate of $0.016667$ Hz (60 s measurement time).
Relevant terms in the analysis are evaluated at 2 kHz, the rate of the available data, and decimated down to $0.016667$ Hz rate, taking into account the information provided by the G group.
\begin{table}[]
    \centering
    \begin{tabular}{c|c|c|c}
       Gas pressure  &  10 mbar\\
        Area of the beam at waist  & $0.51\times0.74\time 10^{-6}$ m$^2$\\
        Mirror transmission & 0.2 ppm\\
        Perimeter of the cavity & 16 m\\
        Kinetic discharge temperature  & 360 K\\
        Photodiode quantum efficiency & 0.5\\
        Trans-impedance amplifier gain  & $1.1\times10^{8}$ V/A\\
        Beat note mean value  & 348.516 Hz\\
        average loss $<\mu>$  &  $6.51280\times 10^{-5}$\\
        Scale factor  & $6.3211125158\times 10^6$ Hz s/rad
    \end{tabular}
    \caption{Data of the G RLG set up.}
    \label{tab:Gparameters}
\end{table}

A set of explanatory variables is evaluated: the laser explanatory variables $\omega_\xi$, $\omega_{ns1}$, $\omega_{ns1} \times \delta\mu$, and the available geodetic terms. 
The LR procedure has been repeated using different schemes. In the first scheme the LR uses the whole set of terms all at once. In the second scheme the LR  has been applied in two steps: the first step evaluates $\omega_s$ using only the laser terms, leading to an initial evaluation of the laser corrections.
In the second step the geodetic terms are used as explanatory variables. When available,  $FGEO$ is added as a known signal and used for check. The different schemes produce always very similar results: in the following,  the first scheme is used, since it is the more conservative one.
 Care has been put to avoid local minima in the regression, which sometimes occurred.
 
\subsection{The data}
The two data sets have 3 full days stored at 2 kHz rate, day 99, 100, and 101, year 2020, and day 121, 122, and 123, year 2022. The two data sets are different and the 2020 one presents more  short duration glitches than the other.
The G data, all the related analysis steps, the employed geophysical models, and environmental monitors are stored on a daily basis in a file at 0.016667 Hz rate. Details of data associated to each column of the published files can be found in the Appendix: in the following, the column numbers will be used in the analysis description. Column 34  provides the measured Earth rotation with the the estimated backscatter correction applied and the theoretical diurnal polar motion and solid Earth tides model subtracted. Since the absolute orientation of the G structure is not known well enough, these latter corrections have to be considered as preliminary. Several tiltmeters are used in order to establish the long term changes of the sensor orientation. These in turn have to be corrected for temperature related effects and from mass attraction derived from a global weather model \cite{kluegel}. In particular the latter comes with several days delay, hence it can not be corrected in realtime. 
The initial value of the ring laser orientation is taken from a local survey, which provides values with a substantial error, much larger than 5 ppm. Typical values are  49.144802 degrees for the latitude and 12.87630 degrees for the longitude. 
While the analysis is based on the methods developed for GINGERINO, there are several differences in the experimental readout system. In GINGERINO the interferogram is the beat note taken at one corner and the monobeams are both measured at the same neighboring corner. For G, the available interferogram is the sum of the two beams at the beam combiner, and the monobeams are each measured at a different corner in order to avoid perturbations from back reflections. In this case the $\omega_\xi$ terms are effective in correcting differences in the mirror transmission, and other differences due to the measurement at different corners. 

\section{Analysis results}
 The beat note $\omega_m$ is evaluated with our method and compared with the one recorded in the published file, contained in column 3. 
 
 There are small differences in the standard deviation, which in the present analysis turns out smaller by $3-4$ $\mu$Hz, probably due to the band-pass filter around the beat note applied before the frequency reconstruction. This is relevant to reduce the impact of the laser systematic terms $\omega_{K1}$  and $\omega_{K2}$ \cite{DiVirgilio2019, DiVirgilio2020}, first and second order expansion, since the second harmonic contribution, which is very difficult to model and subtract, is effectively eliminated in this way. 
 
 Remarkably, in the present data sets irregular signals occur in the separation between one and the subsequent day, in fact to keep constant the number of data for each day some points are missing at the end of each daily file, standard procedure of the Miniseed data retrieval. In the analysis the missing points are replaced with zeroes. 
%
\subsection{Days 99-101, 2020}
We have analysed the days 99, 100 and 101 starting from raw data. Published files, containing one day data, are  extracted from a continuous logging sequence with the help of a MINISEED data format, and, to prevent loss of precision, a set of data in a MATLAB readable format has been used.  Analogously to the procedure used for GINGERINO, data are band-passed around $\omega_m = 348$ Hz, in a $\pm 12$ Hz interval. The relevant parameters are evaluated (AC and DC of the monobeams, relative phase of the mode $\epsilon$, and the beat note $\omega_m$) at 2 kHz rate. Different terms of the analysis, $\omega_{s0}$, $\omega_{\xi}$, and $\omega_{ns1}$ are evaluated using the available information summarized in Table \ref{tab:Gparameters}, including  photodiode gain, parameters of the G RLG (size and gas pressure), the measured average losses $<{\mu}>$. Six $\omega_\xi$ terms are considered, to take into account  errors in the evaluation of $\epsilon$, $I_{S1,2}$, and $PH_{1,2}$ of the monobeams.

The first step of the analysis evaluates $\omega_{s0}$ using Eq.~\ref{approx}. Remarkably, $\omega_{s0}$ values remain rather stable within each single day, but small discontinuities are evident between one day and the other, due to the fact that some points around midnight are missing.

For this reason data around midnight have been eliminated from the analysis: no other cuts have been applied to the data.

\begin{figure}[ht] 
    \centering 
    \includegraphics[scale=0.7]{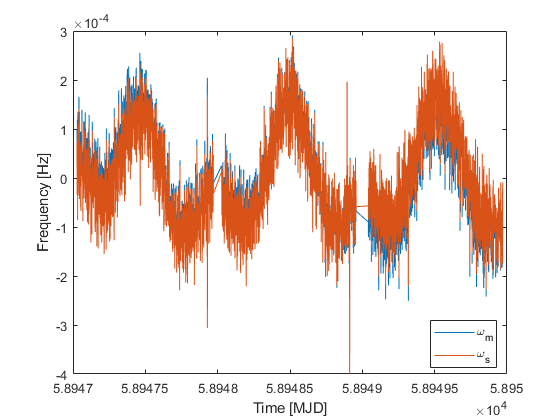}
    \caption{Comparison of  $\omega_s$ (red) and $\omega_m$ (blue) evaluated on the 2020 three days data set: mean values are subtracted in both cases. The 2020 three days data set is considered}
    \label{fig:corr1}
\end{figure}

Figure \ref{fig:corr1} shows $\omega_s$, mean value subtracted, evaluated by the LR procedure for the three days (red solid line), and the discontinuities between different days  are no longer evident. $\omega_m$, mean values subtracted, is also shown (blue solid line): small differences with $\omega_s$ are seen.

\begin{figure}[ht] 
    \centering
    \includegraphics[scale=0.7]{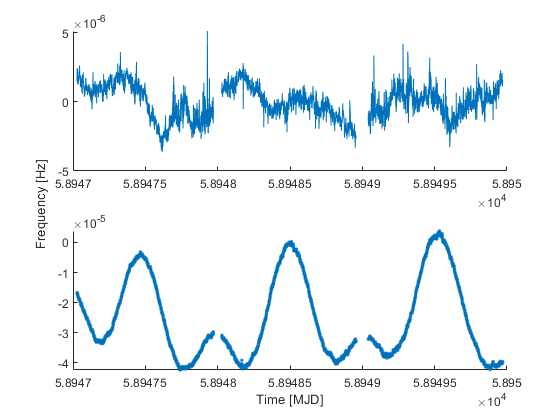}
    \caption{Sum of the laser terms  (top), total geodetic components as evaluated by the procedure (bottom), without including Chandler wobble data. The 2020 three days data set is considered. }
    \label{fig:corr2}
\end{figure}

The top panel of Fig.~\ref{fig:corr2} reports the  $\omega_{ns1}$ corrections evaluated by the LR procedure. The total geodetic components  are plotted in the bottom panel, as evaluated by the LR procedure without including Chandler wobble data.

\subsection{Days 121-123, 2022}

\begin{figure}[ht] 
    \centering
    \includegraphics[scale=0.7]{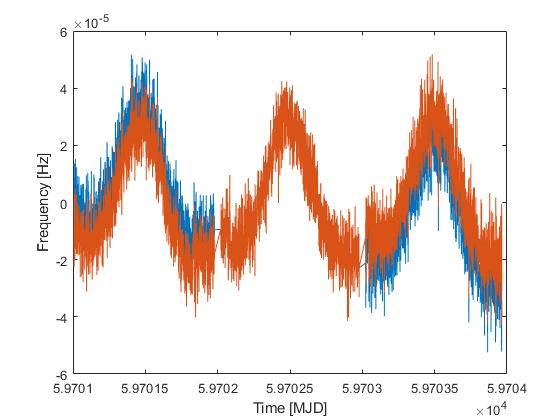}
    \caption{Same of Fig.~\ref{fig:corr1} for the 2022 three days data set.}
    \label{fig:422}
\end{figure}

A second, more recent, set of three days (day 121, 122, and 123, year 2022)  has been also analysed. In this second data set the G RLG was operating under ideal conditions and at roughly 20\% lower beam power. The signals are cleaner, although small discontinuities between different days are still present, but fewer points have been eliminated around midnight to cure the problem. The $\omega_s$ has been evaluated analogously to the first data set. Data indicate that the signal is more stable compared with the previous set of data, accordingly the null shift terms have little impact, but remain to be meaningful for the Allan deviation analysis. Figure \ref{fig:422} and Fig.~\ref{fig:corr1} are very similar with each other. 

\section{Residuals and Allan deviation of the two data sets}
\begin{figure}[ht] 
    \centering
    \includegraphics[scale=0.7]{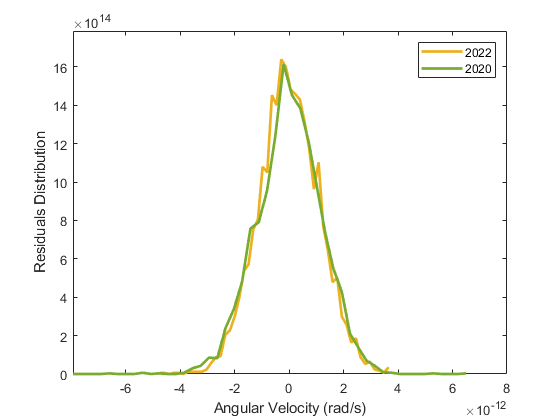}
    \caption{Distribution of the residuals for the two data sets expressed as angular velocity. The standard deviations are $7.73\pm0.05$ $\mu$Hz and $7.14\pm0.05$ $\mu$Hz for the 2020 and 2022 data sets, respectively.}
    \label{fig:res}
\end{figure}

The two data sets exhibit very similar results, however, more points have been eliminated in the 2020 case, probably in response to the higher beam power setting. The amount of data retained for the analysis is $93.4\%$ of the total for the 2020 and $95.8\%$ of the total for the 2022 data set. All cuts had to be applied around midnight. Figure \ref{fig:res} shows the distributions of the residuals for both sets; they are very similar with respect to each other. This indicates that the noise from the laser dynamics has been effectively identified and subtracted by the procedure in both cases.

\begin{figure}[ht] 
    \centering
    \includegraphics[scale=0.5]{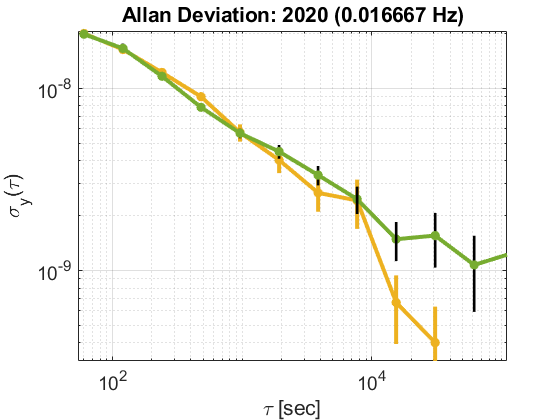}
    \includegraphics[scale=0.5]{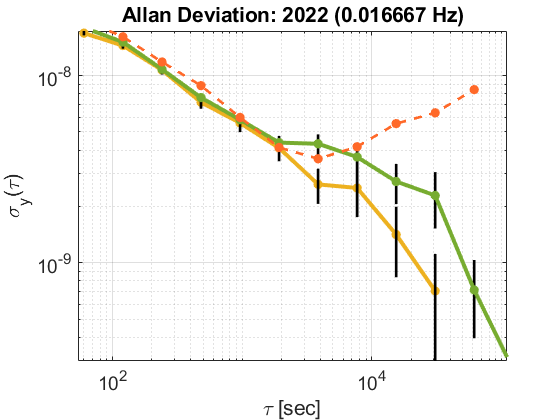}
    \caption{LEFT: 2020, Allan deviation obtained according to the available analyses, for the 3 days (green) and  day 99 (orange). RIGHT: 2022, Allan deviation obtained according to the available analyses, for the 3 days (green) and  day 121 (orange). The red dashed line in the right figure shows the Allan deviation subtracting only the geophysical components from the beat note $\omega_m$.}
    \label{fig:Allan1}
\end{figure}

Figure \ref{fig:Allan1} shows the  Allan\footnote{The function allan of Matlab has been used.} deviation of the residuals for both data sets, sampled at 0.016667 Hz. The green solid lines display results obtained by applying the LR procedure at once on the whole three days data sets. The procedure has been also applied to each single day separately: orange solid lines report the corresponding results (day 99, year 2020, and day 121, year 2022, are considered as examples). Remarkably, plots cross the 1 part in $10^9$ threshold in less than one day of integration time, albeit with a larger standard deviation. 

 The same procedure has been applied to each single day separately and the two panels in Fig.~\ref{fig:Allan1} refer to the Allan deviation of the residuals. Curves obtained in the present analysis are always below those coming out from the standard one, with the exception of one of the days, around $\tau \simeq 8\times 10^3$ s. When $FGEO$ contributions are taken into account, the present analysis leads to even smaller Allan deviation. In any case, plots cross the 1 part in $10^9$ threshold with less than 1 day of integration time.

At the present stage we do not investigate the nature of the residuals,
since the focus is here to investigate whether the Allan deviation goes  below the fundamental physics threshold, i.e., below about 1 part in $10^9$, while keeping the analysis as simple as possible. Moreover, owing to its monolithic structure, G is certainly less prone to those mechanical coupling effects accounted for, in GINGERINO analysis, by the extra term based on the product of the residuals and the tiltmeter signal.

The geophysical components have been subtracted to $\omega_m$ using the LR, without taking into account the laser terms $\omega_\xi$ and $\omega_{ns1}$. With the 2022 data set, definitely less affected by laser dynamics, the Allan deviation is a factor 2 worse at $10^4$ seconds compared to Fig.~\ref{fig:Allan1} RIGHT, where the Allan shown with the dashed red line is super-imposed.
With the 2020 data set, the result is even higher, indicating that, despite  G  is based on an extremely rigid optical cavity, 
it is necessary to identify and subtract the laser systematic terms in order to reach, and go beyond, the 1 part in $10^9$ precision level of the Earth rotation rate measurement.

\section{Conclusions}
The analysis based on the model developed for GINGERINO has been extended to the G RLG, rewriting the mathematical relationships with the main aim to account for the characteristic values of the G apparatus.
In this way, corrections to the laser systematics have been subtracted  in a deterministic way following the model based on the stationary solution of the RLG equations and the loss $\bar{\mu}$ of the optical cavity determined by statistical means. The signals of geodetic origin  have been always subtracted assuming linear relationships to minimize the residuals via linear regression.
 For the full set of three days of 2020, the Allan deviation drops below one part in $10^9$ in less than 1 day of integration time.

 The data point close to midnight are eliminated, to overcome a problem in the data retrieval subroutine and the separate analysis of each single day exhibits Allan deviations going  below 1 part in $10^9$ more rapidly.

 The analysis has been repeated with a second, and more recent, set of three days in June 2022, where the ring laser was operated at lower power. In this case, the monobeam signals are cleaner, and the cavity appears to be more stable, accordingly the effects of $\omega_{ns1}$ are smaller, but still remain significant for the lower Allan deviation outcome.
  
 
 G is based on a monolithic structure in ZERODUR, a very low thermal expansion material, with mirrors optically contacted to the structure. For this reason the cavity is extremely stable, nevertheless tiny laser systematic effects are present and it is necessary to subtract them in order to improve the performance above the 1 part in $10^9$ level. Certainly laser systematics cancellation is more relevant for  RLGs based on a HL design,  as GINGERINO and ROMY, since in this case cavity losses, the quantity effectively ruling laser systematics,  is affected by variations of the mirror distances and their relative alignment. 
 

\appendix
\section{The G daily file with data and results}
On a daily basis, G data sampled at 0.0166667 Hz rate and relevant results are published in a file.
Contents of each column are listed in Table \ref{tab:column}.

\begin{table}
    \centering
    \begin{tabular}{|c|c|c|}
    \hline
    \hline
    Column & Quantity & Units\\
\hline
\hline
 1& Epoch& [MJD]\\
 \hline
  2& Epoch& [day]\\
 \hline
  3& Sagnac (single tone extractor)& [Hz]\\
  \hline
4&Sagnac rms & [Hz] \\
\hline
5& Phase between SW-port and Sagnac beam combiner port& [rad]\\
\hline
6& Phase between SE-port and Sagnac beam combiner port& [rad]\\
\hline
 7& SW-AC &[V]\\
 \hline
 8&SW-DC& [V]\\
 \hline
 9& SW-AC/DC & -\\
 \hline
10& SE-AC& [V]\\
\hline
 11& SE-DC& [V]\\
 \hline
 12& SE-AC/DC&-\\
 \hline
 13& SW-SE-Phase& [rad]\\
 \hline
 14& estimated Sagnac correction factor&-\\
\hline
 15& estimated Sagnac correction value& [mHz]\\
\hline
 16& Sagnac BS-corrected& [Hz]\\
\hline
 17& SE-DDS-Amplitude AC-level for driving LED& [V]\\
 \hline
 18& SW-DDS-Amplitude AC-level for driving LED& [V]\\
 \hline
 19& SW-Phase to DDS driving LED& [rad]\\
 \hline
20& SE-Phase to DDS driving LED& [rad]\\
\hline
21&Sagnac geophys. model contribution subtracted& [Hz]\\
& (Oppolzer-Terms $\&$ Tilt-NS deformation effect)& \\
\hline
22& Sagnac of SE-monobeam estimation (Single Tone extractor)& [Hz]\\
\hline
 23& Sagnac of SW-monobeam estimation (Single Tone extractor)& [Hz]\\
\hline
 24& Geophysical Model: Oppolzer terms NS& [rad]\\
 \hline
 25& Geophysical Model: Oppolzer terms EW& [rad]\\
 \hline
  26& Geophysical Model: Theoretical tilt NS (attraction part)& [rad]\\
\hline
 27& Geophysical Model: Theoretical tilt NS (deformation part)& [rad]\\
\hline
 28& Observed tilt NS& [rad]\\
 \hline 
 29& Geophysical Model: Theoretical tilt EW (attraction part)& [rad]\\
 \hline
 30& Geophysical Model: Theoretical tilt EW (deformation part)& [rad]\\
 \hline
  31& Observed tilt EW& [rad]\\
  \hline
 32& Geophysical Model: Sagnac geophys. model contribution& [$\mu$Hz]\\
 & (Oppolzer-Terms $\&$ Tilt-NS deformation effect)& \\
\hline
 33& Sagnac geophys. model reduced & [Hz]\\
 &(Oppolzer-Terms $\&$ Tilt-NS deformation effect)& \\
 \hline
34& Sagnac geophys. model and BS applied& [Hz]\\
\hline
35&Pressure vessel barometric pressure & [hPa]\\
&(Paroscientific at pressure vessel supply)& \\
\hline
36& Pressure vessel temperature& [$^o$C]\\
\hline
37& Pressure vessel humidity& [$\%$rH]\\
\hline
38& Ringlaser room barometric pressure & [hPa]\\
&(Vaisala outside pressure vessel)& \\
\hline
39& Ringlaser room temperature& [$^o$C]\\
\hline
40& Ringlaser room humidity& [$\%$rH]\\
\hline
41& Control room temperature& [$^o$C]\\
\hline
42& Control room humidity& [$\%$rH]\\
\hline
43& Plasma brightness& [V]\\
\hline
\hline  
\end{tabular}
\caption{Content of columns in the G data and results file.}
\label{tab:column}
\end{table}

\end{document}